\documentclass{aip-cp}

\usepackage[numbers]{natbib}
\usepackage{rotating}
\usepackage{graphicx}
\usepackage{url}
\usepackage{aas_macros_mod}
\usepackage{soul}
\usepackage{xcolor}

\usepackage{hyperref}

\newcommand{\pos}[1]{\href{http://pos.sissa.it/cgi-bin/reader/contribution.cgi?id=#1}{\tt #1}}

\newcommand{\Swift}{\textit{Swift}}
\newcommand{\FERMI}{\textit{Fermi}}

\begin{document}

% Title portion
\title{The Exceptional Flare of Mrk\,501 in 2014 \\ Combined Observations with H.E.S.S. and FACT}

%\author[aff1]{G.\,Cologna\corref{cor1}}
\author[]{G.\,Cologna$^{\mathrm{1,a}}$}
%\eaddress{gabriele.cologna@gmail.com}
%\author[aff2]{N.\,Chakraborty}%\corref{cor1}}
\author[]{N.\,Chakraborty$^{\mathrm{2,b}}$}
%\eaddress{cnachi@mpi-hd.mpg.de}
\author[aff3]{A.\,Jacholkowska}
%\author[aff4]{M.\,Lorentz}%\corref{cor1}}
\author[]{M.\,Lorentz$^{\mathrm{4,c}}$}
%\eaddress{matthias.lorentz@cea.fr}
\author[aff1]{M.\,Mohamed}
\author[aff3,aff5]{C.\,Perennes}
\author[aff6]{C.\,Romoli}
\author[aff1]{S.\,J.\,Wagner}
%\author[aff7,aff1]{A.\,Wierzcholska for the H.E.S.S. Collaboration}
\author[]{A.\,Wierzcholska$^{7,1}$ for the H.E.S.S. Collaboration}
%\author[aff8,aff9]{D.\,Dorner for the FACT Collaboration}
\author[]{D.\,Dorner$^{\mathrm{8,9,d}}$ for the FACT Collaboration}
%\author[]{D.\,Dorner$^{8,9}$ for the FACT Collaboration}
%\eaddress{dorner@astro.uni-wuerzburg.de}
\author[aff10,aff1]{O.\,Kurtanidze}

\affil[aff1]{Landessternwarte, Universit\"at Heidelberg, K\"onigstuhl 12, D 69117 Heidelberg, Germany}
\affil[aff2]{Max-Planck-Institut f\"ur Kernphysik, P.O. Box 103980, D 69029 Heidelberg, Germany}
\affil[aff3]{LPNHE, Universit\'e Pierre et Marie Curie Paris 6, Universit\'e Denis Diderot Paris 7, CNRS/IN2P3, 4 Place Jussieu, F-75252, Paris Cedex 5, France}
\affil[aff4]{18 DSM/Irfu, CEA Saclay, F-91191 Gif-Sur-Yvette Cedex, France}
\affil[aff5]{LUTH, Observatoire de Paris, CNRS, Universit\'e Paris Diderot, 5 Place Jules Janssen, 92190 Meudon, France}
\affil[aff6]{Dublin Institute for Advanced Studies, 31 Fitzwilliam Place, Dublin 2, Ireland}
\affil[aff7]{Insitute of Nuclear Physics, Polish Academy of Sciences, Radzikowskiego 152, PL-31-342 Krakow, Poland}
\affil[aff8]{Universit\"at W\"urzburg, D-97074 W\"urzburg, Germany}
\affil[aff9]{Universit\"at Erlangen-N\"urnberg, Physikalisches Institut, Erwin-Rommel-Str. 1, 91058 Erlangen, Germany}
\affil[aff10]{Abastumani Observatory, Mt. Kanobili, 0301 Abastumani, Georgia}

%\corresp[cor1]{Corresponding author: contact.hess@hess-experiment.eu\\ $^{\mathrm{b})}$dorner@astro.uni-wuerzburg.de}
\corresp[cor1]{gabriele.cologna@gmail.com, $^{\mathrm{b})}$cnachi@mpi-hd.mpg.de, $^{\mathrm{c})}$matthias.lorentz@cea.fr,\\ $^{\mathrm{d})}$dorner@astro.uni-wuerzburg.de}

\maketitle

\begin{abstract}
The BL\,Lac type object Mrk\,501 was observed at very high energies (E\,$>$\,100\,GeV) in 2014 with the upgraded H.E.S.S. (High Energy Stereoscopic System) phase 2 array. The data collected with the central 28\,m telescope allow for a broader energy range extending to lower energies when compared to the one obtained with the four small telescopes alone. A strong flaring event with a flux level comparable to the 1997 historical maximum has been detected as a consequence of target of opportunity observations triggered by alerts from the FACT collaboration. The First G-APD Cherenkov Telescope (FACT) is continuously monitoring bright blazars at TeV energies providing important pre- and post-flare information. For the first time, the data and lightcurves from H.E.S.S. and FACT are compared. These contemporaneous observations allow for a better characterization of the source emission. In a multiwavelength context, more precise correlation studies between VHE and lower energies are possible thanks to the dense sampling of the FACT observations. The hard intrinsic spectrum detected by H.E.S.S. during the flare allows the derivation of strong constraints on the scale of Lorentz invariance violation via the non-detection of EBL opacity modifications and from time-of-flight studies.
\end{abstract}

%\keywords{BL Lac Objects, Mrk\,501, EBL}

\section{INTRODUCTION}

Mrk\,501 ($z\sim$\,0.034) is a blazar belonging to the subclass of the high frequency peaked BL\,Lac objects (HBL). Its emission is strongly variable in all energy bands from radio up to very high energies (VHE, E$>$100\,GeV), where it was discovered as the second extragalactic source in 1995 \cite{Whipple1995_1996ApJ...456L..83Q}. Its synchrotron and inverse Compton (IC) peaks are known to move towards higher energies during high flux states. In particular, the synchrotron peak can shift up to two orders of magnitude \cite{Pian1997_1998ApJ...492L..17P}. As a consequence, a flux dependent spectral hardening has been observed in both X-rays and in the GeV-TeV band \cite[\textit{e.g. }][]{Pian1997_1998ApJ...492L..17P, CAT_1997_1999A&A...350...17D, FERMI_mwl2008_2009_2011ApJ...727..129A, cologna_2015arXiv150904458C}.

Mrk\,501 has been extensively monitored over the years, often in multiwavelength (MWL) campaigns in order to characterize its spectral energy distribution (SED). In 1997, a prolonged high state was detected \cite[\textit{e.g. }][]{CAT_1997_1999A&A...350...17D, Whipple1997_1997ApJ...487L.143C, Whipple1997_1998ApJ...501L..17S, HEGRA_1997A&A...327L...5A, HEGRA1997_spectrum_1999A&A...349...11A, HEGRA1997_temp_char_1999A&A...342...69A} and the historical highest flux was measured. Comparable high fluxes have been observed in short events on a few occasions, as in 2005 \cite{MAGIC_2005_2007ApJ...669..862A}, 2012 \cite{MAGIC_ICRC_2013} and 2014 \cite{cologna_2015arXiv150904458C}. During flares, the VHE flux variability can be as short as a few minutes \cite{MAGIC_2005_2007ApJ...669..862A, nachi_2015arXiv150904893C}.

\section{OBSERVATIONS AND ANALYSIS}

\textbf{VHE} In 2014, the continuous monitoring of Mrk\,501 with FACT \cite{FACT_design_2013JInst...8P6008A} lead to the detection of several high state events with fluxes above a few Crab units (c.u.). In every one of these cases, flare alerts were issued, two of which triggered Target of Opportunity (ToO) observations with H.E.S.S.. A flare with flux comparable to the 1997 historical maximum was detected by both instruments on the night of June 23-24 2014.
Mrk\,501 was observed in wobble mode with the full HESS\,II array on several nights between June 19-25 and on July 29-30, and a total of 21 runs (observations with a typical exposure of 28 minutes) were collected. 18 runs pass the standard quality cuts, for a total livetime of 7.7\,h (7.1\,h when corrected for acceptance). The mean zenith angle is 63.8$^\circ$ causing the energy threshold of the observations to be very high, above 1\,TeV. In \cite{cologna_2015arXiv150904458C}, only events detected with the CT1-4 telescopes were extracted from the observations. Here, the information coming from all five telescopes are considered. The Model analysis \cite{deNaurois2009APh....32..231D} was used to reduce the stereo data. \textit{Very Loose} cuts were used in order to obtain the lowest possible energy threshold. This choice does not raise any issue regarding background contamination since the background level is very low.
% Being the background contamination (determined with the \textit{Reflected Region Background} method \cite{Berge2007A&A...466.1219B}) very low, \textit{Very Loose} cuts were used in order to obtain the lowest possible energy threshold. 
An excess of almost 1900 photons is detected with high significance. Spectral analyses have been carried out on the total dataset and on the flare (four runs in one night) and low state (all runs in the remaining five nights) separately using the forward-folding technique \cite{Piron_2001A&A...374..895P}. All results have been cross-checked using a different software and an independent calibration.
%The FACT monitoring provides valuable pre- and post-flare information. Thus, FACT observations between May 24 and August 4 are presented in this work alongside the H.E.S.S. observations. A total of 317\,h in 68 nights have been collected. After applying the quality cuts, 150\,h in 44 nights remained, with an average zenith angle of ppp$^\circ$. 
The FACT monitoring provides valuable pre- and post-flare information. Thus, FACT observations between May 24 and August 4 are presented in this work alongside the H.E.S.S. observations. A total of 204 h in 63 nights have been collected. For the conversion to c.u., data with zenith angle smaller than 45$^{\circ}$ and only limited amount of moon light have been selected, leaving 144.5 hours of data in 59 nights. Rejecting bad weather data as described in \cite{dorner_2013arXiv1311.0478D} and excluding all data suffering from calima (the effect of calima on IACT data is described in \cite{dorner_2009A&A...493..721D}), 104 hours of data from 41 nights remain with an average zenith angle of 21.6$^{\circ}$. In the night of the flare, 4.1 hours of data were recorded. To allow for fast flare alerts, FACT has a quick-look analysis (described in \cite{dorner_2015arXiv150202582D}) providing results with low latency.

% In this proceeding, the focus will be on the 2014 observations, which comprise the strongest flare ever detected with H.E.S.S. Afforded by the unique data set, the VHE variability is investigated. It is at energy scales higher than reported before and the aim is to improve and extend constraints on the source emission mechanisms. Also, comparisons with the 2006 flare of PKS 2155-304 \cite{2010A&A...520A..83H} are instructive in looking for general features in the temporal structure of the emission and therefore for potential principles underlying the physics of jet emission. 

\textbf{HE} The Large Area Telescope (LAT) onboard the \FERMI\, satellite provides the data at high energies (HE, 100\,MeV\,$<$\,E\,$<$\,100\,GeV). The observations collected in a four months period centered on the VHE flare have been analysed with the ScienceTool software package version \verb|v9r33p0|. Events belonging to the 'Source' class within 15$^\circ$ from the position of Mrk\,501 were selected, and cuts on the zenith angle (100$^\circ$), rocking angle (52$^\circ$) and distance from the Sun (5$^\circ$) were applied. The instrument response functions \verb|P7REP_SOURCE_V15| and the standard Isotropic and Galactic diffuse emission background models iso\_source\_v05.txt and \verb|gll_iem_v05 rev1.fit| were used for the binned maximum-likelihood spectral analysis. Mrk\,501 was not sufficiently bright for sufficiently long time in order to derive short time-scale spectra or lightcurves with time bins shorter than one week.

\textbf{X-rays} X-ray data collected with \Swift-XRT in WT mode between May 23 and August 6, 2014 (obsIDs: 00035023026-00035023078) have been analysed for this work using the HEASoft software package v.\,6.16 (\url{http://heasarc.gsfc.nasa.gov/docs/software/lheasoft}) with CALDB v.\,20140120. The events were cleaned and calibrated using \verb|xrtpipeline| and the data in the energy range 0.3-10\,keV with grades 0-2 were analysed. For the spectral fit, the events were grouped in bins with at least 30 counts using the \verb|grappha| tool and were fit using XSPEC v.\,12.8.2 with a single power-law model and Galactic hydrogen absorption fixed to $n_H=1.58\cdot 10^{20}$\,cm$^{-2}$ \cite{Kalberla05_gal_abs}. The lightcurve flux values (Figure\,\ref{fig:longterm}) were calculated integrating the spectra of single snapshots between 2 and 10\,keV. The relation between fluxes and spectral indices shows a clear harder-when-brighter behaviour, as expected from the literature (e.g. \cite{Pian1997_1998ApJ...492L..17P}). Most of the X-ray and FACT observations are simultaneous or at least contemporaneous, while X-ray and H.E.S.S. observations are typically offset by $\sim$\,90\,minutes in the nights with coverage of both instruments. 

\textbf{Optical} Observations in optical R band have been collected with the 70\,cm telescope of the Abastunami Observatory (Georgia), which is equipped with an Apogee 6E camera. The data has been analysed with the Daophot II reduction software using an aperture diameter of 10$^{''}$ \cite{omar_1999bmtm.proc...25K}. Also in this case, nights of contemporaneous data with H.E.S.S. and FACT exist. Notably, strictly simultaneous observations have been taken during the flare peak on June 23, 2014.

\section{MWL LIGHTCURVES}

\begin{figure*}[t]
  \centering
  \includegraphics[width=1.0\columnwidth]{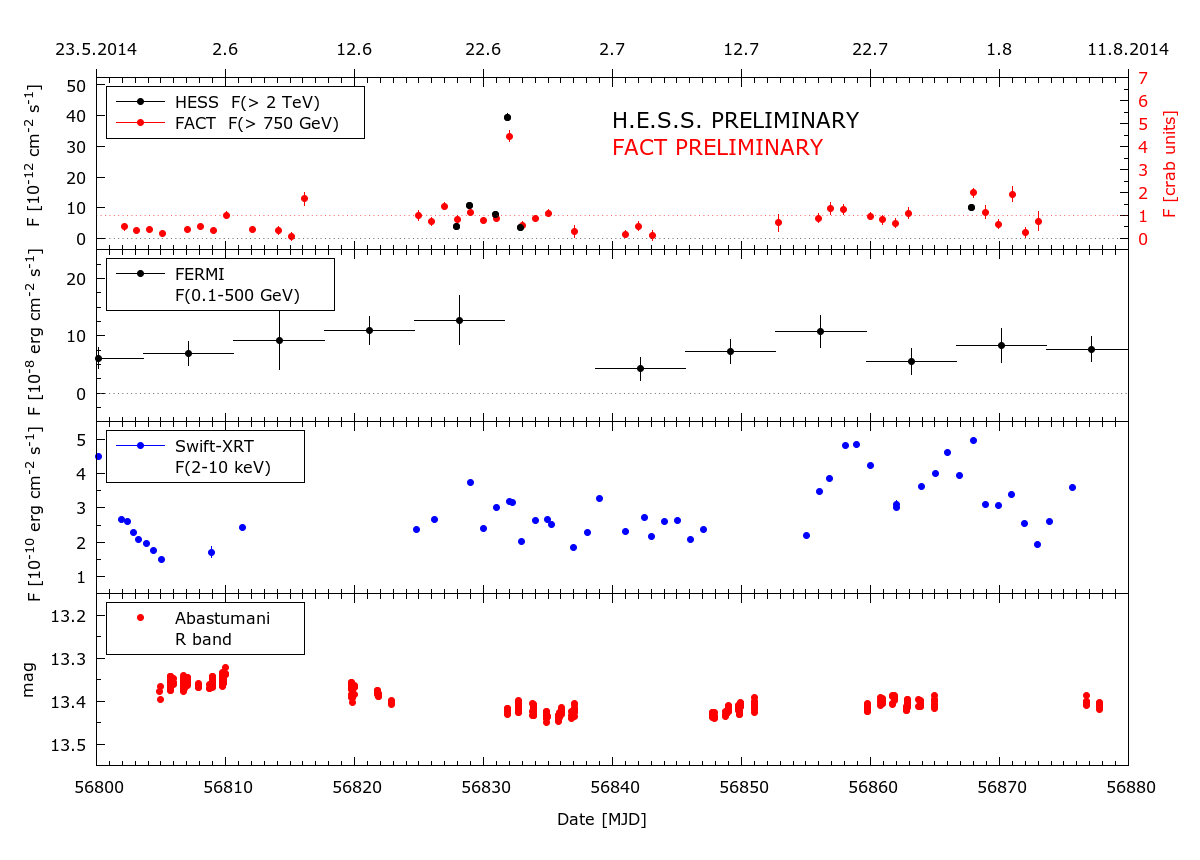}
  \caption{From \textit{top} to \textit{bottom}. First panel: VHE night-wise lightcurves from H.E.S.S. (black) and FACT (red). The fluxes are estimated above different energy thresholds: 2\,TeV and 750\,GeV, respectively. The FACT fluxes are provided in c.u. only. The flare is clearly visible at MJD 56831-32. The five nights of "low" state in the H.E.S.S. dataset represent a state of moderate activity. Second panel: \FERMI-LAT weekly lightcurve. One point is missing due to too low exposure caused by technical problems on the spacecraft and by pointed-mode observations on another source. Third panel: 2-10\,keV \Swift-XRT X-ray lightcurve showing high variability. Fourth panel: optical lightcurve from the Abastunami Observatory. These are observed magnitudes, not corrected for the host galaxy contribution. No particularly high flux values or strong variability are detected in correspondence of the simultaneously observed TeV flare. The spread of the values within one night represents the uncertainty of the observations.} 
  \label{fig:longterm}
\end{figure*}

%The complementarity of FACT and H.E.S.S. was well exploited for the H.E.S.S. observations of the flare in June 2014 triggered by FACT. FACT issued triggers including on the nights of 19th and 23rd which initiated and prolonged the target-of opportunity observations with the full array of H.E.S.S. 
The regular monitoring of Mrk\,501 by FACT ensures dense sampling, which is of primary importance in providing valuable pre- and post-flare information, as in the current case. 
%As a result, we have recorded the TeV behaviour leading up to the $\sim 5$ Crab peak on the night of 24th June and the subsequent decline of flux. 
The high sensitivity of H.E.S.S. on the other hand allows for minute dissection of the flare itself, probing timescales down to a few minutes.
% Thus, this collaborative effort between H.E.S.S. and FACT was able to track the rise and fall of this record flare (with flux levels comparable to the historical maximum of 1997). 

In Figure\,\ref{fig:longterm}, the multiwavelength lightcurves spanning 80 days around the June 23 flare are shown: VHE data from H.E.S.S. (black, E\,$>$\,2\,TeV) and FACT (red, E\,$>$\,750\,GeV) in the top panel, HE data from \FERMI-LAT (0.1-500\,GeV) in the second panel, X-ray data from Swift (2-10\,keV) in the third panel and optical R-band data from Abastumani in the bottom panel. As can be seen in the top panel, the H.E.S.S. and FACT observations cover the flux rise from less than 1\,c.u. before June 19 to $\sim$\,5\,c.u. of the peak emission on the night of June 23, as well as the following sharp decline to sub c.u. levels.

\begin{figure*}[t]
  \centering
  \includegraphics[width=0.32\columnwidth]{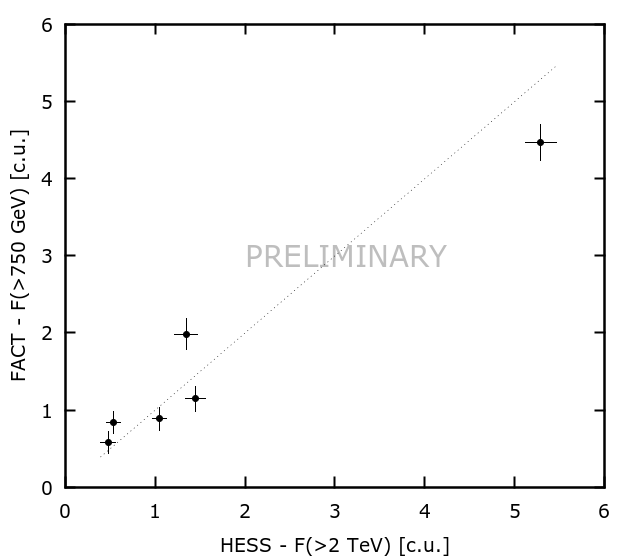}
  \includegraphics[width=0.32\columnwidth]{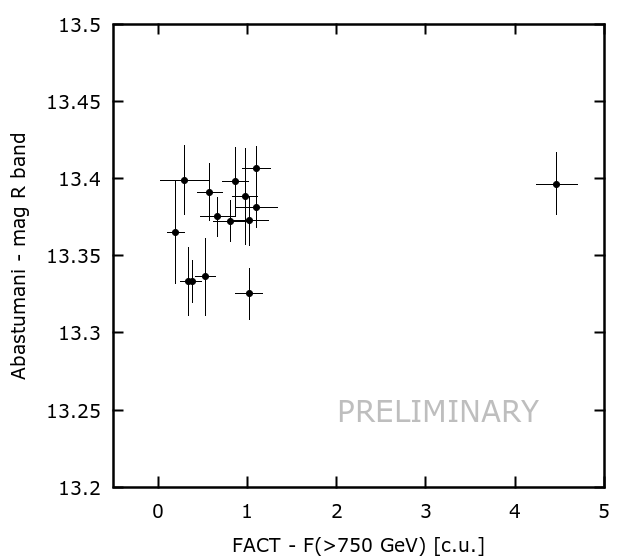}
  \includegraphics[width=0.32\columnwidth]{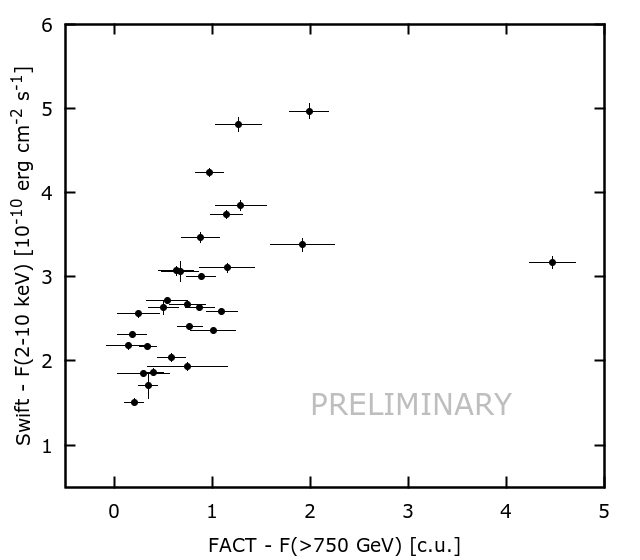}
  \caption{Comparison of the nightly average fluxes in different energy bands. \textit{Left}: a simple linear correlation is recovered as expected between H.E.S.S. and FACT. \textit{Middle}: The optical emission level seems to have no relation with the VHE one. \textit{Right}: A positive correlation can be seen between X-rays and VHE. It is violated by the right most point, suggesting differences in the physical parameters leading to the $\gamma$-ray emission during the flare with respect to the rest of the campaign.}
  \label{fig:correlation}
\end{figure*}

Despite the different energy thresholds and not complete simultaneity, the average nightly fluxes recorded with H.E.S.S. and FACT in the same nights correlate well (Figure\,\ref{fig:correlation}, \textit{left}). In this period, there was no significant activity either in the optical or at HE. The comparison of the average nightly optical magnitudes with the fluxes measured by FACT (Figure\,\ref{fig:correlation}, \textit{middle}) shows no correlation. In particular, the optical emission is stable on the average value during the strictly simultaneous observations of the peak of the flare. The harder-when-brighter behaviour detected in X-rays can in principle explain these findings, since it requires the variability at lower energy to have a lower amplitude. On the other hand, it is known that the source shows activity also in optical. This suggests that at least two zones or mechanisms are necessary to explain the variability in VHE and in optical.

%Figure\,\ref{fig:correlation}, \textit{right} The X-ray observations by Swift during this period showed nightly variability, but do not show any clear correlation with TeV.  

\begin{figure*}[t]
  \centering
  \includegraphics[width=0.6\columnwidth]{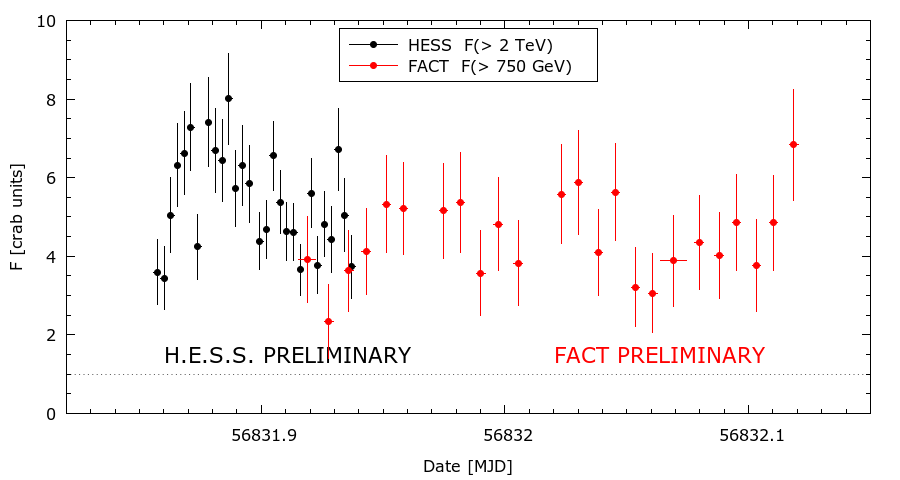}
  \caption{Zoom in of the H.E.S.S. (black) and FACT (red) lightcurves during the June 23 flaring event. The binning is 4 and 10 minutes for the two instruments, respectively. While minute-scale variability is hidden by the low sensitivity of FACT, flux doubling times below 10 minutes are detected by H.E.S.S. . Note the different energy range used.}
  \label{fig:shorttime}
\end{figure*}

Figure\,\ref{fig:shorttime} shows the flaring state on June 23 on minute level time scale. The H.E.S.S. and FACT lightcurves have bins of 4 and 10 minutes, respectively. 
There is evidence of fast variations on timescales of a few minutes at TeV energies. Doubling times below 10 minutes are obtained from the H.E.S.S. data. This favours a lepton induced emission at VHE. 
%It is interesting to note, however, that a possible quadratic correlation expected between X-ray and TeV fluxes, and indicative of synchrotron self-Compton (SSC) emission, is violated by the right-most point in Figure\,\ref{fig:correlation}, \textit{right}. Again, this suggests that other processes are at play during the 2014 flaring event.
%It is interesting to note, however, that the right-most point in Figure\,\ref{fig:correlation}, \textit{right}, violates the correlation shown by the other points. This suggests that the source state during the 2014 flare is different compared to the rest of the campaign in terms of the physical parameters of the emitting region or of the particles responsible for the $\gamma$-ray emission.
It is interesting to note, however, that the correlation expected between X-ray and TeV fluxes, indicative of synchrotron self-Compton (SSC) emission, is violated by the right-most point in Figure\,\ref{fig:correlation}, \textit{right}. Again, this suggests that the source state during the 2014 flare is different compared to the rest of the campaign in terms of the physical parameters of the emitting region or of the particles responsible for the $\gamma$-ray emission.

\section{TeV SPECTRA}

%\begin{figure*}[t]
%  \centering
%  \includegraphics[width=0.5\columnwidth]{mrk501_2014_ECPL_all_flare_low_stereo_prod8_sq.png}
%  \caption{Observed H.E.S.S. spectra for the three subset described in the text. The visible curvature is due only to EBL absorption.}
%  \label{fig:spectra}
%\end{figure*}

In \cite{CAT_1997_1999A&A...350...17D} and \cite{MAGIC_2005_2007ApJ...669..862A}, flux dependent spectral variability has been reported. In those publications, it was determined mostly at energies below 2\,TeV. In the present work, the high energy threshold above 1\,TeV and the very high maximal energy reached allow for a flux dependent spectral analysis at energies complementary to those of previous publications. Besides the total 2014 dataset, also the flare and the low state nights have been analysed separately.

Power law (PL) spectral fits provide very poor results, in line with the previous findings by, \textit{e.g.}, \cite{CAT_1997_1999A&A...350...17D, HEGRA1997_spectrum_1999A&A...349...11A, MAGIC_2005_2007ApJ...669..862A}, and with the consideration that the intrinsic emitted spectrum must have been absorbed through the interaction with the extragalactic background light (EBL). In fact, despite Mrk\,501 being a close source, a significant optical depth for EBL absorption $\tau\gtrsim$1 is expected at the highest energies observed from the source. Unsurprisingly, the curved power law (CPL) and the exponential cut-off power law (ECPL) spectral shapes fit the data satisfactorily. However, being the optical depth significant, the intrinsic spectra will differ from the observed ones. For this reason, fits including EBL absorption from \cite{Franceschini_2008A&A...487..837F} have been performed.

The spectra are best described by an EBL-absorbed PL. The photon indices are $1.91\pm0.04_{stat}\pm0.2_{sys}$ for the flaring state, and $2.32\pm0.06_{stat}\pm0.2_{sys}$ for the low state. The respective normalizations at 3.5\,TeV are $13.9\pm0.3$ and $2.7\pm0.1$ 10$^{-12}$cm$^{-2}$s$^{-1}$TeV$^{-1}$. There is no evidence of intrinsic curvature nor intrinsic cut-off up to 25\,TeV. These results show how a flux dependent spectral variability exists for energies extending significantly beyond 10\,TeV, up to 20\,TeV on a nightly time interval. The fact that the intrinsic spectrum is best described by a simple hard PL during the 2014 flare also indicates that there is no sign of being in the Klein-Nishina regime even at these high energies.

\section{LORENTZ INVARIANCE VIOLATION STUDIES}

The 2014 flare data set with an energy coverage extending significantly up to 20\,TeV has naturally been used to put constraints on Lorentz invariance violation (LIV).

%\subsection{Modified dispersion relation}
Deviations from Lorentz invariance could appear at an energy scale beyond our current grasp. A generic approach to introduce such effects consists in adding effective terms in the dispersion relation of particles, \textit{i.e.} for photons
\begin{equation}
%E_\gamma^2 = p_\gamma^2 \pm E_\gamma^2  \left( \frac{E_\gamma}{E_{\mathrm{LIV}}} \right)^n
%\Longleftrightarrow 
E_\gamma^2 = p_\gamma^2 \left[1 \pm \left(\frac{E_\gamma}{E_{\mathrm{LIV}}} \right)^n\right],
\label{eq:LIV_Dispersion}
\end{equation}
where $E_{\mathrm{LIV}}$ is the hypothetical energy scale at which Lorentz symmetry could stop being exact, and $n$ is the leading order of the perturbation. In some approaches to quantum gravity, $E_{\mathrm{LIV}}$ is expected to be of the order of Planck energy  
$\mathrm{E}_{\mathrm{Planck}} = \sqrt{\hbar c^5 /G} \simeq 1.22 \times 10^{28}$\,eV \cite{amelino_2013LRR....16....5A}.

Energetic astrophysical $\gamma$-rays can be used to constrain $E_{\mathrm{LIV}}$ as Equation\,\ref{eq:LIV_Dispersion} would lead to non-negligible observational effects. The most direct approach in searching for LIV is to look for energy-dependent time delays in the arrival time of $\gamma$-rays. Constraints obtained with this approach using the 2014 flare data set are described in \cite{cologna_2015arXiv150904458C}. These results are comparable with the best time-of-flight constraints for the linear case ($n=1$) derived from AGN observations \cite{liv_pks2155_2011APh....34..738H}, and are currently the best time-of-flight constraints for the quadratic case ($n=2$).

%\subsection{LIV-modified EBL absorption}
An attractive alternative possibility to test Lorentz invariance takes advantage of the fact that the presence of LIV would affect the kinematics for the $e^+e^-$ pair production of TeV $\gamma$-rays on the EBL, resulting in a modified opacity. In the subluminal case we consider here (minus sign in Equation\,\ref{eq:LIV_Dispersion}), the threshold energy for the pair creation process is increased. This would reduce the EBL absorption and result in an excess of transparency for $\gamma$-rays at the highest energies with a specific spectral signature \citep[see \textit{e.g.}][]{Jacob_2008PhRvD..78l4010J}.

The spectral characteristics of Mrk 501 during flares and its low redshift allow for the detection of the most energetic extragalactic $\gamma$-rays, making this source ideal to investigate LIV through spectral studies,
% as it has already pointed out in studies using the 1997 outburst \cite{Stecker_2001APh....16...97S,biteau_2015ApJ...812...60B,Tavecchio_2016A&A...585A..25T}, or in prospective studies on the sensitivity of CTA to LIV \cite{Fairbairn_2014JCAP...06..005F}.
as has already pointed out by, \textit{e.g.}, \cite{Stecker_2001APh....16...97S,biteau_2015ApJ...812...60B,Tavecchio_2016A&A...585A..25T} using the 1997 outburst, or by \cite{Fairbairn_2014JCAP...06..005F} discussing the sensitivity of CTA to LIV.

%\subsection{Method and results}
Optical depths using the EBL model of \cite{Franceschini_2008A&A...487..837F} are computed considering modifications due to LIV for the subluminal case. A log-likelihood spectral fit is performed on the H.E.S.S. 2014 flare dataset assuming an intrinsic power-law and letting the spectral index and the normalization free to vary. Values of $E_{\mathrm{LIV}}$ are scanned logarithmically in the range of interest for linear and quadratic scenarios. More details can be found in \citep{matthias_2016arXiv160608600L}. As the data show no evidence for a high-energy upturn, LIV-free optical depth values are preferred and log-likelihood profiles reach plateaus corresponding to the standard case. This allows to compute exclusion limits on $E_{\mathrm{LIV}}$, as summarized in Table\,\ref{tab:LimitsTable}.

\begin{table*}[b]
\centering
\caption{Exclusion limits on $E_{\mathrm{LIV}}$ at different confidence levels obtained from the spectral analysis of the 2014 flare data set.}
\label{tab:LimitsTable}
\tabcolsep7pt\begin{tabular}{c|ccc}
  \hline \\[-0.8em]
       & 2$\sigma$                                                      & 3$\sigma$                                                     & 5$\sigma$     \\[0.2em] 
  \hline \\[-0.8em]
  n=1  & $3.3\times 10^{28}$\,eV ($2.67\times\mathrm{E}_{\mathrm{Planck}}$) & $2.6\times 10^{28}$\,eV ($2.13\times\mathrm{E}_{\mathrm{Planck}}$) & $1.7\times 10^{28}$\,eV ($1.37\times\mathrm{E}_{\mathrm{Planck}}$)     \\[0.2em] 
  n=2  & $8.7\times 10^{20}$\,eV \hspace*{\fill}                         & $7.8\times 10^{20}$\,eV \hspace*{\fill}                        & $6.3\times 10^{20}$\,eV \hspace*{\fill}      \\[0.2em] 
  \hline 
\end{tabular}
\end{table*}

For the linear case, the 5\,$\sigma$ lower limit is at $1.7\times 10^{28}$\,eV ($1.37 \times \mathrm{E}_{\mathrm{Planck}}$). It is the best limit obtained using an AGN and confirms the result obtained with time delays using GRB\,090510 \cite{liv_grb_2013PhRvD..87l2001V}: the standard photon dispersion relation holds up to the Planck energy scale in the case of linear perturbations.
For the quadratic case, the 5\,$\sigma$ lower limit is at $6.3\times 10^{20}$ eV, which increases the current limits obtained with both AGNs and GRBs. These strong constraints naturally come from the exceptional spectrum of the 2014 flare data-set where the power law intrinsic emission extends up to 20 TeV with no sign of deviations from standard EBL absorption.

The same LIV analysis using the EBL model of \cite{Dominguez_2011MNRAS.410.2556D} leads to very similar exclusion limits, showing that the results are not significantly affected by the choice of a specific EBL model. We have considered LIV affecting only photons \cite[as in ][]{Tavecchio_2016A&A...585A..25T,Fairbairn_2014JCAP...06..005F} and not electrons since the constraints on LIV for electrons are very stringent \cite{Liberati_LIV_2003Natur.424.1019J}.
% \footnote{Relaxing this condition and considering LIV affecting electrons and photons equally amounts to a rescaling of $E_{\mathrm{LIV}}$ by $(1-2^{-n})$.}.
Relaxing this condition and considering LIV affecting electrons and photons equally amounts to a rescaling of $E_{\mathrm{LIV}}$ by $(1-2^{-n})$ \citep{Jacob_2008PhRvD..78l4010J}.

% If not already mentioned, say that the spectrum has been xchecked with independent calibration and analysis chains, the general conclusion of this paragraph being that these results should not be dominated by systematics.

% Acknowledgement
\section{ACKNOWLEDGMENTS}
\footnotesize{The H.E.S.S. Collaboration acknowledges: The support of the Namibian authorities and of the University of Namibia in facilitating the construction and operation of H.E.S.S. is gratefully acknowledged, as is the support by the German Ministry for Education and Research (BMBF), the Max Planck Society, the German Research Foundation (DFG), the French Ministry for Research, the CNRS-IN2P3 and the Astroparticle Interdisciplinary Programme of the CNRS, the U.K. Science and Technology Facilities Council (STFC), the IPNP of the Charles University, the Czech Science Foundation, the Polish Ministry of Science and Higher Education, the South African Department of Science and Technology and National Research Foundation, and by the University of Namibia. We appreciate the excellent work of the technical support staff in Berlin, Durham, Hamburg, Heidelberg, Palaiseau, Paris, Saclay, and in Namibia in the construction and operation of the equipment.
The FACT Collaboration acknowledges: The important contributions from ETH Zurich grants ETH-10.08-2 and ETH-27.12-1 as well as the funding by the German BMBF (Verbundforschung Astro- und Energy Physics; the Laboratory Astroteilchenphysik) and HAP (Helmoltz Alliance for Astro-particle Physics) are gratefully acknowledged. We are thankful for the very valuable contributions from E. Lorenz, D. Renker and G. Viertel during the early phase of the project. We thank the Instituto de Astrofisica de Canarias allowing us to operate the telescope at the Observatorio del Roque de los Muchachos in La Palma, the Max-Planck-Institut für Physik for providing us with the mount of the ormer HEGRA CT3 telescope, and the MAGIC collaboration for their support.
G.C. and M.M. are members of the International Max Planck Research School for Astronomy and Cosmic Physics at the University of Heidelberg (IMPRS-HD) and the Heidelberg Graduate School of Fundamental Physics (HGSFP).
N.C. acknowledges support from Alexander von Humboldt foundation.
The Abastumani Observatory team acknowledges financial support by the Shota Rustaveli NSF through project FR/577/6-320/13.}

% References

%\nocite{*}
\bibliographystyle{aipnum-cp}%
\bibliography{biblio_gamma2016}%

\end{document}